\documentclass[12pt]{spieman}  % 12pt font required by SPIE;
\usepackage{amsmath,amsfonts,amssymb,aas_macros}
\usepackage{graphicx}
\usepackage{setspace}
\usepackage{tocloft}
\title{Impact of crosshatch patterns in H2RGs on high precision radial velocity measurements: Exploration of measurement and mitigation paths with HPF}
\author[a,b,*]{Joe P. Ninan}
\author[a,b]{Suvrath Mahadevan}
\author[a,b]{Gudmundur Stefansson}
\author[c]{Chad Bender}
\author[d]{Arpita Roy}
\author[c]{Kyle F. Kaplan}
\author[e,f]{Connor Fredrick}
\author[e,f]{Andrew J. Metcalf}
\author[a,b]{Andrew Monson}
\author[g]{Ryan Terrien}
\author[a,b]{Lawrence W. Ramsey}
\author[e,f]{Scott A. Diddams}

\affil[a]{The Pennsylvania State University, Department of Astronomy \& Astrophysics, University Park, PA 16803, USA}
\affil[b]{The Pennsylvania State University, Center for Exoplanets \& Habitable Worlds, University Park, PA 16803, USA}
\affil[c]{The University of Arizona, Department of Astronomy and Steward Observatory, Tucson, AZ, USA}
\affil[d]{The California Institute of Technology, Department of Astronomy, Pasadena, CA, USA}
\affil[e]{Time and Frequency Division, National Institute of Standards and Technology, 325 Broadway, Boulder, CO 80305, USA}
\affil[f]{Department of Physics, University of Colorado, 2000 Colorado Avenue, Boulder, CO 80309, USA.}
\affil[g]{Carleton College, Department of Physics \& Astronomy, Northfield, MN, USA}

\cftpagenumbersoff{figure}
\cftpagenumbersoff{table} 
\begin{document} 
\maketitle

\begin{abstract}
Teledyne's H2RG detector images suffer from cross-hatch like patterns which arises from sub-pixel quantum efficiency (QE) variation. In this paper we present our measurements of this sub-pixel QE variation in the Habitable-Zone Planet Finder's H2RG detector. We present a simple model to estimate the impact of sub-pixel QE variations on the radial velocity, and how a first order correction can be implemented to correct for the artifact in the spectrum. We also present how the HPF's future upgraded laser frequency comb will enable us to implement this correction. 
\end{abstract}

% Include a list of up to six keywords after the abstract

% Include email contact information for corresponding author
{\noindent \footnotesize\textbf{*}Joe P. Ninan,  \linkable{jpn23@psu.edu} }

\begin{spacing}{2}   % use double spacing for rest of manuscript

\section{Introduction}
\label{sect:intro}  % \label{} allows reference to this section
The Habitable-Zone Planet Finder (HPF) is a stabilized fiber-fed near-infrared (0.81 to 1.28 $\mu$m) ultra-stable precision radial velocity (RV) spectrometer commissioned on the 10 m Hobby-Eberly Telescope (HET) at McDonald Observatory, with the scientific goal of discovering and confirming low-mass planets around M dwarf stars. HPF uses a 1.7 $\mu$m cutoff H2RG array (Hawaii-2RG HgCdTe 2048x2048) cooled to 120 Kelvin as the detector \cite{Mahadevan2014}.
We have demonstrated intrinsic calibration precision as low as 6 cm/s and the measurement of differential stellar RVs at the 1.53 m/s level over months-long time scales, which is unprecedented in the near infrared (NIR) wavelength region \cite{Metcalf2019}. Nevertheless, there are still avenues for improvement in the precision of NIR RVs. 

NIR HxRG detectors suffer from various artifacts compared to optical CCDs, which need to be corrected for precision spectroscopic measurements.
One of the artifacts which affects the sub-pixel quantum efficiency (QE) of pixels in HxRG are the sub-pixel crosshatch patterns. These are believed to be intra-pixel QE variations due to lattice defects in the HgCdTe crystal layer\cite{Hardy2008,Shapiro2018,Crouzet2018}. Little work exists on measuring the dimensions of these structures and their impact on RV measurements under the traditional flat correction scheme. In this manuscript, first we present our simple step model in Section \ref{sect:impactonRV} to study the impact of a sub-pixel QE defect  on the spectrum. We also derive a simple model to quantify the impact on the RV estimate as a function of the model parameters and the density of crosshatch in H2RG. In Section \ref{sect:FlatMeasure}, we present a method to estimate the characteristic width, angle, and the QE inside the defect from 2D flat images. Based on these models, we present our proposed correction algorithm in Section \ref{sect:CorrAlgo}, and how an upgraded HPF's laser frequency comb (LFC) will enable us to estimate the required coefficients for the proposed correction algorithm. We conclude the manuscript in Section \ref{sect:conclusion}.

\section{Impact on precision RV}
\label{sect:impactonRV}
\subsection{Modeling of intra-pixel QE variation due to crosshatch pattern in H2RG}
\begin{figure}
\begin{center}
\begin{tabular}{c}
\includegraphics[width=0.4\textwidth]{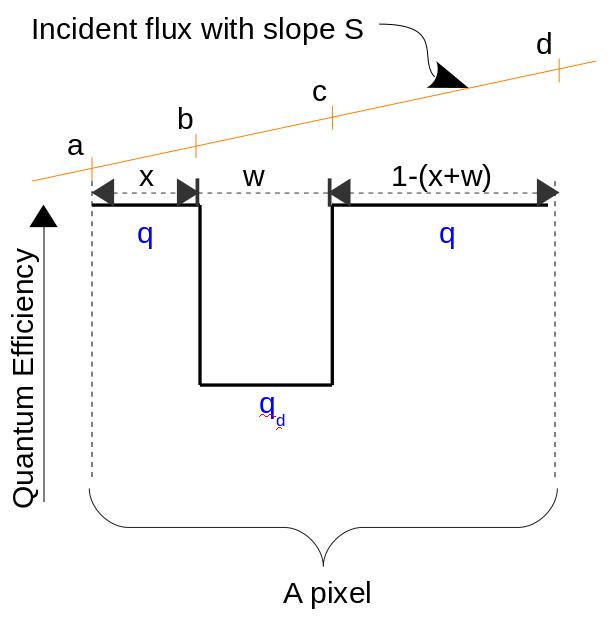}
\end{tabular}
\end{center}
\caption 
{ \label{fig:QEcross}
Our simple one dimensional step function model of intra-pixel QE difference due to a crosshatch in the pixel. The width of the defect is $w$, and the QE inside and outside the defect is $q_d$ and $q$ respectively.} 
\end{figure}

Consider the cross section of a pixel with a QE defect shown in Figure \ref{fig:QEcross}. Let $q$ be the QE outside the defect, and $q_d$ be the QE inside the defect which has a width $w$. Let $x$ be the fractional position inside the pixel where the defect starts.

\subsection{Average gain correction using flat illumination}
The average pixel gain is traditionally corrected by first illuminating the detector with a flat light source. The pixel to pixel relative gain factor is obtained from the measured deviation in pixel counts from neighbouring pixels. Let $C_{Flat}$ be the counts we would have obtained from a pixel with no defect. In our defective pixel model, the measured count is given by $C_{FlatMeasured} = C_{Flat}[wq_d + (1-w)q]$. Hence, the standard flat correction recipe to convert measured counts to average pixel gain corrected count is to divide by $[wq_d + (1-w)q]$, i.e., $C_{Flat} = \frac{C_{FlatMeasured}}{[wq_d + (1-w)q]}$.   

\subsection{Uncorrected counts due to non-flat illumination}
When a defective pixel which has intra-pixel QE variation is illuminated with a non-flat illumination source, the average QE correction estimated from flat illumination source in previous section is not valid. This will lead to an over correction or under correction by the flat fielding process. For modelling this error term, consider an illuminating light source with slope $S$ in pixel coordinates as shown in Figure \ref{fig:QEcross}. Let $a$, $b$, $c$ and $d$ be the incident flux values at $0$, $x$, $x+w$ and $1$ positions inside the pixel. Let $C_s$ be the net counts from the pixel we would have obtained after flat correction if there were no defects in the pixel. i.e., $C_s = \frac{a+d}{2}\times 1$. Since $S$ is the slope of the spectrum inside the pixel, we can parametrise $a = C_s -S/2$, $d = C_s +S/2$, $b =a+xS$, and $c=a+(x+w)S$.
The measured counts in the defective pixel before flat correction is given by $C_{sMeasured} = \frac{a+b}{2}qx + \frac{b+c}{2}q_d w + \frac{c+d}{2}q(1-(x+w))$.
Substituting $a$, $b$, $c$, and $d$, and simplifying, we obtain $C_{sMeasured} = C_s[w q_d + (1-w)q] + Sw(q-q_d)[\frac{1-w}{2} -x]$.
The average flat correction by dividing $[wq_d + (1-w)q]$ outlined in previous subsection, yields the flat corrected count as, 
\begin{equation}
\label{eq:CHcorrection}
C_{sMeasuredFlatCorrected} = C_s + \frac{Sw(q-q_d)[\frac{1-w}{2} -x]}{[wq_d + (1-w)q]} \,.
\end{equation}

Equation \ref{eq:CHcorrection} shows the effect of crosshatch pattern is an additive term to the final counts from the pixel. This term is a linear function of the slope of the spectrum inside a pixel. As expected, due to symmetry, when the crosshatch is in the middle of the pixel (i.e., $x = \frac{1-w}{2}$ the second term vanishes. Also, under the limiting case of QE inside the defect being same as outside (i.e., $q_d = q$), or when $w = 0$ the second term vanishes and $C_{sMeasuredFlatCorrected} = C_s$.

Typically in spectrographs, the spectrum is spread over a number of pixels in the cross-dispersion direction (width of the slit/fiber). Let $N$ be the number of pixels the spectrum is spread in the cross-dispersion direction. After 2D to 1D sum extraction of the spectrum (we assume sum extraction here for easy of explanation, rather than optimal extraction which is what we use in reality), the total counts in the $i^{th}$ pixel of the 1D spectrum will be, 

\begin{equation}
\label{eq:CHcorrectionN}
A(i) = \sum\limits_{j}^N C_{s_j} + \sum\limits_{j}^N  \frac{Sw(q-q_d)[\frac{1-w}{2} -x(j)]}{[wq_d + (1-w)q]} \,,
\end{equation}

where $x$, $q_d$ are dependent on $j$ ($j$ is the index of the pixel in cross-dispersion direction).

Let the true spectrum which we would have obtained if there was no defect in the $i^{th}$ pixel be $A_o(i)$. Then $A_o(i) = \sum\limits_{j}^N C_{s_j}$.
If the spectrograph's cross dispersion slit profile is top hat shaped like in the case of HPF, we can simplify $C_{s_j} = A_o(i)/N$, and the slope in pixel space $S = \frac{1}{N}\frac{dA_o}{d\lambda}\Delta\lambda$, where $\Delta\lambda$ is the wavelength dispersion per pixel \footnote{For a general cross dispersion profile, one has to keep this summation as it is. However, it does not change the additive nature of the equation nor inferences in the subsequent sections of this manuscript.}.

\subsection{RV error introduced by the crosshatch pattern}
The fundamental equation to calculate the radial velocity from the change in flux of a spectrum at pixel $i$ is given by the formula \cite{Bouchy2001}
\begin{equation}
\label{eq:RVofpixel}
\frac{\delta V(i)}{c} = \frac{A(i) - A_o(i)}{\lambda(i) [\frac{dA_o}{d\lambda}]_i} \,.
\end{equation}

Substituting Equation \ref{eq:CHcorrectionN} in Equation \ref{eq:RVofpixel}, we obtain the expression for the spurious RV  induced by the crosshatch,
\begin{equation}
\label{eq:RVduetodefect}
\frac{\delta V(i)}{c} = \frac{\Delta\lambda}{\lambda(i)} \frac{1}{N} \sum\limits_{j}^N  \frac{w(q-q_d)[\frac{1-w}{2} -x(j)]}{[wq_d + (1-w)q]} =  \frac{\Delta\lambda}{\lambda(i)} \xi,
\end{equation}
where
\begin{equation}
\label{eq:Xi}
\xi = \frac{1}{N} \sum\limits_{j}^N  \frac{w(q-q_d)[\frac{1-w}{2} -x(j)]}{[wq_d + (1-w)q]} \, .
\end{equation}

We note an important feature in the expected RV noise from a single pixel is that this quantity is independent of the slope of the spectrum!
For pixels containing crosshatch, $x(j)$ is a straight line whose angle and position can be measured from the flat images as detailed in Section \ref{sect:FlatMeasure}.

\subsection{RV impact on HPF}

\begin{figure}
\begin{center}
\begin{tabular}{c}
\includegraphics[width=0.6\textwidth]{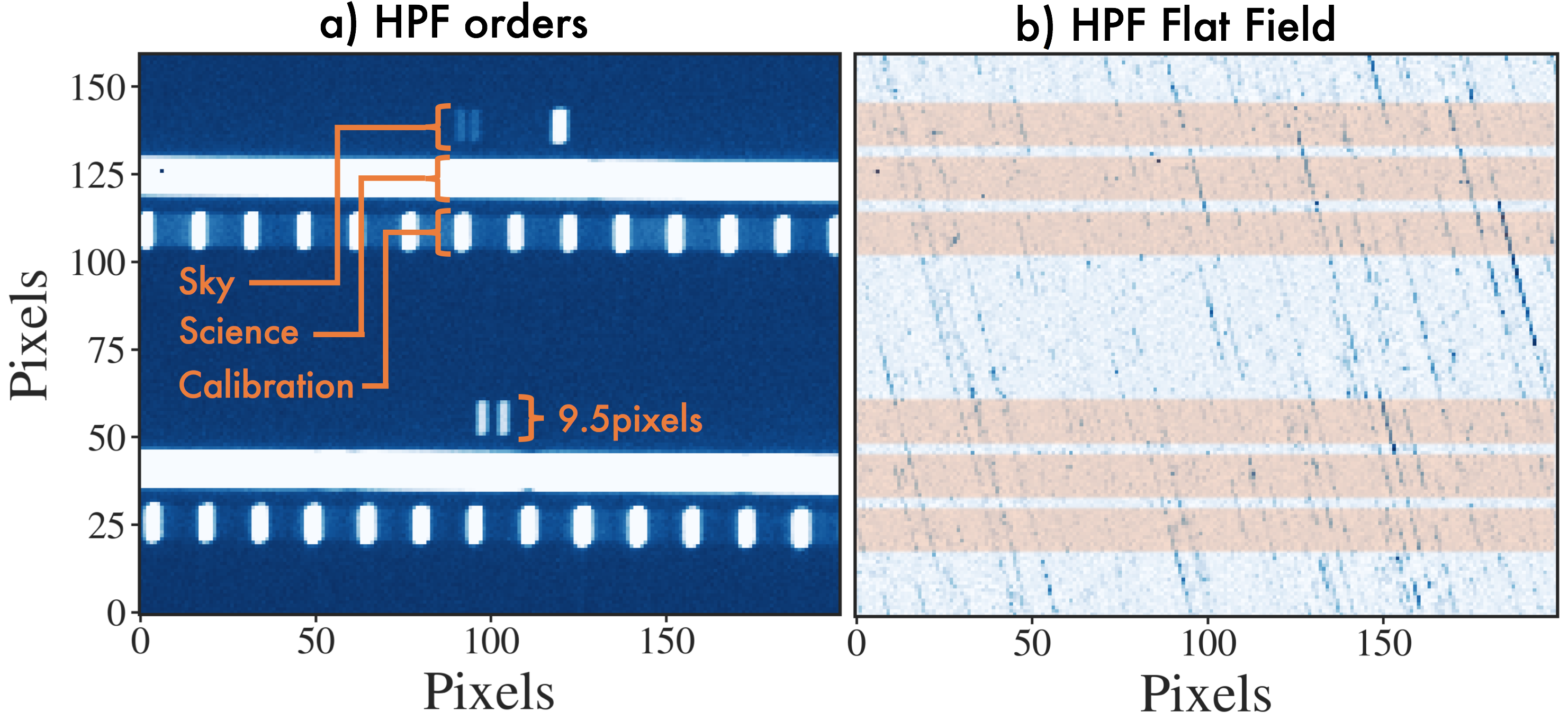} 
\end{tabular}
\end{center}
\caption 
{ \label{fig:HPFtrace}
a) A zoom in on-sky science image showing only two orders showing the HPF science, sky and calibration order traces. b) The same as a) but during a filtered flat field exposure to illustrate the size of the cross hatch pattern with order traces overlaid.} 
\end{figure}

\begin{figure}
\begin{center}
\begin{tabular}{c c}
\includegraphics[width=0.45\textwidth]{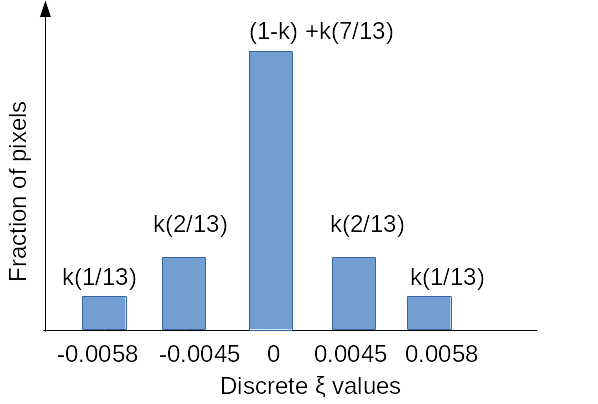} & \includegraphics[width=0.5\textwidth]{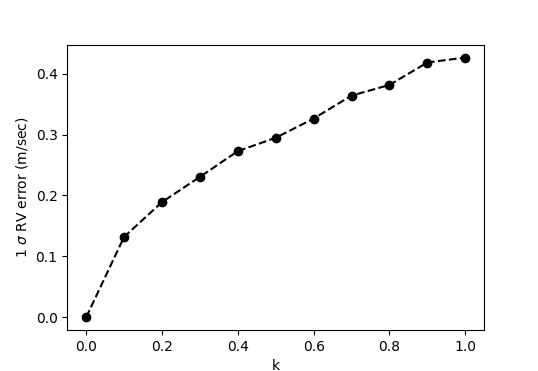}\\
(a) & (b) \\
\end{tabular}
\end{center}
\caption 
{ \label{fig:HPFRVprobabilityCrossHatchRV}
a) The probabilistic histogram of pixel level discretised $\xi$ values causing the crosshatch induced RV errors in HPF. $k$ is the fraction of pixels in a spectrum affected by crosshatch defects. b) 1 sigma RV error as a function of $k$ for a typical mid M-dwarf star spectrum in a typical order of HPF due to crosshatch pattern.} 
\end{figure}

The $q_d$ inside the crosshatch of HPF's H2RG varies across the detector. A typical value in HPF for $q_d/q$ is $0.54$, width $w=1/3.78$, and the slope of the line defining $x(i)$ is $14.8^o$ (see Section \ref{sect:FlatMeasure} for more details on these measurements). This  $14.8^o$ implies a continuous line of sub-pixel crosshatch defect crosses 3.8 pixel rows while moving from one column pixel to the next. HPF was designed to image the fiber slit across 2.5$\times$9.5 pixels as shown in Figure \ref{fig:HPFtrace}. To enable this averaging, the rectangular slit was rotated and carefully aligned vertically along the pixel columns. Due to the symmetry of the $[\frac{1-w}{2} -x]$ term in Equation \ref{eq:RVduetodefect}, RV error will cancel out if all the 3.8 pixels affected by a crosshatch defect are fully inside the 9.5 pixel slit image of HPF.  However, when the crosshatch pattern appears near the edge pixels of the slit they do not cancel out, resulting in a residual error. By discretising the positions of crosshatch pattern on a 10 pixel slit column, we can approximate HPF's pixel level RV errors to five discrete $\xi$ values. Figure \ref{fig:HPFRVprobabilityCrossHatchRV}(a) shows the probabilistic histogram of these pixel level $\xi$ values for an order where $k$ fraction of the column pixels in that order are affected by crosshatch defects.

The net stellar RV measured by the least-squares technique\cite{Anglada2012,Zechmeister2018} is equivalent to optimal weighted average of the pixel level RV values given by the Equation \ref{eq:RVduetodefect}. The optimal weights are given by Equation 8 in Bouchy et. al (2001)\cite{Bouchy2001}, which is proportional to $\lambda^2(i) \frac{dA}{d\lambda}^2$. Using HPF's dispersion solution, along with the spectrum of a typical M-dwarf star (specifically, Barnard's Star), we calculated the optimal weights and thereby the weighed average of pixel level RVs drawn from the probability distribution of $\xi$ shown in Figure \ref{fig:HPFRVprobabilityCrossHatchRV} a). Figure \ref{fig:HPFRVprobabilityCrossHatchRV} b) shows the typical 1 sigma error due to crosshatch defects in the RV estimated for this M-dwarf star from a single order (2040 pixel columns), as a function of $k$, where $k$ is the fraction of pixels in the order affected by crosshatch defects.

\subsection{Caveats of the model}
The major caveat to our model is that we are treating the crosshatch pattern originating from crystal defects as pure sub-pixel low QE regions. We are ignoring all the other possible artifacts in the pixel behaviour due to its impact on electron mobility, diffusion, etc. For the sake of simplicity we are also assuming a step function model for the sub-pixel QE difference. We also made the assumption that the  spectrum after convolution with instrumental PSF is smooth enough to be locally approximated by a line with slope $S$ inside a pixel. To cancel out the crosshatch effect inside an extended slit like in case of HPF, the $q_d$ should not vary. This is only partially true in certain areas of HPF detector. When $q_d$ varies rapidly at small pixel length scales the noise in RV will be even higher than suggested by Figure \ref{fig:HPFRVprobabilityCrossHatchRV} b). The result in Figure \ref{fig:HPFRVprobabilityCrossHatchRV} b) is sensitive to the $q_d$, hence they should be treated only as a typical order of magnitude estimate of the RV error due to the crosshatch defects. The discretization of the $\xi$ for the calculation also makes this result an approximation. In this calculation, we also ignored the effects due to correlation in the presence or absence of crosshatch across nearby column pixels.  Despite these caveats, this model can be easily extended to any sub-pixel QE defects of any detector.

\section{Measuring intra-pixel structure from the flat image}
\label{sect:FlatMeasure}
\begin{figure}
\begin{center}
\begin{tabular}{c c c}
\includegraphics[width=0.31\textwidth]{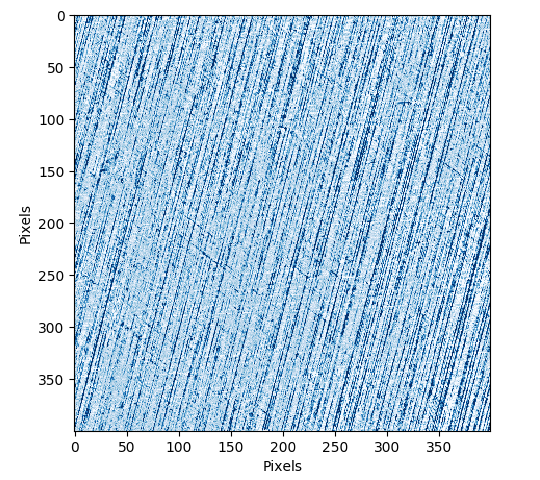} & \includegraphics[width=0.37\textwidth]{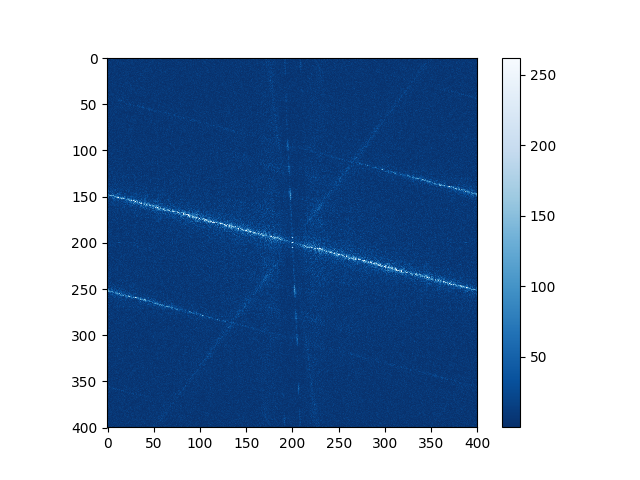}&
\includegraphics[width=0.16\textwidth]{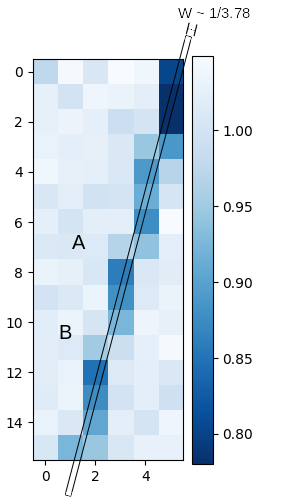} \\
(a) & (b) & (c) \\
\end{tabular}
\end{center}
\caption 
{ \label{fig:crosshatchimage}
a) A sample crosshatch pattern region in HPF. b) 2D Fourier power spectrum of the region showing the angle of the crosshatches, as well as the power extending all the way to Nyquist sampling hinting the sub-structure nature of cross-hatches. c) Zoomed image of a typical 14.8 degree crosshatch QE variation pattern. Labels A and B mark two column crossover points of the subpixel crosshatch defect. A best fitted rectangular sub-pixel crosshatch with a width of 1/3.78 pixels is also overlayed on the pixels.} 
\end{figure}

A uniformly illuminated flat image can be used to estimated the parameters of the model we developed in the previous Section. Figure \ref{fig:crosshatchimage} a) shows a small region of the HPF H2RG detector after high pass filtering of a smoothly illuminated flat data. Ideally, we should also divide out the $e^-/adu$ gain of each pixel to make a cleaner effective QE map of pixels. The angle of the crosshatch pattern can be measured precisely from the 2D power spectrum of this image (Figure \ref{fig:crosshatchimage} b)). The angle was measured to be 14.8 degrees for HPF’s detector. The power in this 2D Fourier power spectrum at Nyquist frequency also shows the intra-pixel structures are indeed under-sampled as expected by the pixel grid. 

\subsection{Defect's width $w$ and QE $q_d$}
Figure \ref{fig:crosshatchimage} c) shows a zoomed image of a typical 14.8 degree crosshatch QE variation pattern in the HPF detector. The defect moves 3.8 rows before it jumps to next column pixel. 

%\begin{figure}
%\begin{center}
%\begin{tabular}{c}
%\includegraphics[width=0.3\textwidth]{CrossHatchZoomInHPF_611-627_1592-1598_AB.png} 
%\end{tabular}
%\end{center}
%\caption 
%{ \label{fig:crosshatchZoomAB}
%Zoomed image of a typical 14.8 degree crosshatch QE variation pattern. Labels A and B mark two column %crossover points of the subpixel crosshatch defect.} 
%\end{figure} 

To aid the discussion, two points of column crossovers are labelled as A and B in the Figure \ref{fig:crosshatchimage} c). Between A and B, the defect is fully contained inside a pixel boundary, and the pixel averaged relative QE of the middle three pixels (with respect to the outside region) is $\sim$88 \%. i.e., a net QE drop of $\sim$12 \%. The region labelled A, has the defect moving from column 3 to 4 within a single row. The sum of the QE drop in both those crossover pixels combined is $\sim$11 \%. In the region labelled as B, the defect is moving from column 2 to 3 across two rows. The sum of the QE drop in the crossover pixels combined is $\sim$10 \% for both the rows. Hence, at least in this small patch of the crosshatch the variation in $q_d$ is small. 

If the width of the defect is very narrow, at 14.8 degrees angle, cross-hatch will always take less than one pixel row to crossover. If the width is large, it will take multiple rows to crossover columns. Thus the shortest and longest crossover length scale (in units of rows) from one column to adjacent column constrains the width of the defect. Based on the 1 and 2 rows crossover width, we can estimate the width of the crosshatch defect ($w$) to be $\sim$1/3.78 of a pixel. i.e., $\sim$5 $\mu m$ in Figure \ref{fig:crosshatchimage} c). This geometrically constrained width of the defect is shown by the overlay.

Substituting $w = 1/3.78$ in the formula for average 88 \% QE (when the defect is fully inside a pixel) from Section \ref{sect:impactonRV};  $[wq_d + (1-w)q] = 0.88 q$, we obtain $q_d/q \sim 0.55$. i.e., the QE inside the defect of $\sim$5 $\mu m$ width is $\sim$55 \% of the region outside the defect. 

\subsection{Caveats of this method}
The QE ($q_d$) inside the defect varies across the defect as one can see from the flat images (Figure \ref{fig:crosshatchimage} a)). So $q_d$ is not a single number and the value derived in the previous Section is just a typical value in HPF's H2RG detector. 
The absolute gain differences of each neighboring pixel’s amplifier limits the accuracy of this QE variation analysis. Hence we have to make an accurate pixel-by-pixel $e^-/adu$ gain-map using
individual pixel's photon transfer curve, and then divide that out from the flat image shown here to improve the analysis.
And lastly, the relative neighboring pixel QE analysis outlined here is difficult in the regions with
high density of crosshatch patterns.

\section{Correction Algorithm}
\label{sect:CorrAlgo}
The vertical rectangular fiber slit of HPF enables us to reduce the 2D intra-pixel inhomogeneity into a simpler 1D problem. In this Section, we outline the rational behind the proposed 1 D spectrum correction algorithm for HPF.

The optical cross dispersion profile of the HPF's trace is locally a well defined shape since the flux contribution from nearby wavelengths is constant across the rectangular slit profile\footnote{Note that this is not generally true for spectrographs fed by bare, non-sliced, and non-rectangular fiber.}. Let's denote this cross dispersion profile across 10 pixels as a 10 dimensional normalized vector, $P$. 
Let $f$ be the scalar quantity which represents the total flux at any given pixel column in the trace. Then the
profile at that column in trace is given by the vector $fP$.
Let's denote the effective gain and QE correction (for a flat spectral source) at each 10 pixels inside the profile by the vector $G$. The sum-extracted 1D spectrum is then given by the dot product of these vectors $= G\cdot fP = f G\cdot P$. This separability of $f$ and $G \cdot P$ at each column in a trace of HPF due to the rectangular slit is the key aspect of HPF which enables us to use this 1 D algorithm.
For a non-flat spectral source like a star ($f_s$), based on Equation \ref{eq:CHcorrection}, we expect an additive term proportional to the slope of the spectrum, i.e., the measured sum-extracted spectrum will be $f_s G\cdot P + c \frac{df_s}{d\lambda}$, where the proportionality constant $c$ encodes all the averaged effective $w$ and $q_d$ values.

To correct for the crosshatch pattern error, we propose the following steps:

Step 1: The first step is to divide the 1D extracted spectrum using the 1D extracted spectrum of a normalized flat illumination source. This step removes the $G\cdot P$ term from the flux resulting in the output to be 
\begin{equation}
\label{eq:effectiveresidue}
f_s + \frac{c}{G\cdot P} \frac{df_s}{d\lambda} \,.
\end{equation}

Step 2: The slope of the spectrum $\frac{df_s}{d\lambda}$ is either calculated iteratively, or from a multi epoch average template spectrum which (due to barycentric/instrumental shift) is not affected by the non-defective pixel.

Step 3: Slope is multiplied with $ c/(G\cdot P)$ to obtain the additive correction which needs to be subtracted out from the extracted spectrum after Step 1 (Equation \ref{eq:effectiveresidue}) to recover $f_s$. The following subsection explains how the required coefficient $ c/(G\cdot P)$ is calculated for each column pixel in the trace of the spectrum. 

\subsection{Calculation of the correction coefficient}
For estimating the proportionately constant of the correction term due to the slope of the spectrum (i.e., $ c/(G\cdot P)$ in Equation \ref{eq:effectiveresidue}), we need a spectral source with known spectral slope and flux. By shifting the spectrum in wavelength or frequency, across each column pixel, we can measure the excess flux as a function of the slope. In theory, a late type star observed under a range of barycentric shift will enable the calibration of at least few regions of the spectrum.
For robust estimation of the proportionality coefficient, one needs a spectral source with maximum slope allowed by the optical PSF of the instrument. This calibration could be performed with a tunable Laser Frequency Comb (LFC) as is further discussed below.

\subsection{Using Laser Frequency Comb for measuring intra-pixel QE}
A sub-pixel tuneable LFC enables us to scan the sharpest instrumental profile of HPF across a pixel yielding maximum leverage on estimating intra-pixel QE variations. The large change in slope of the spectrum when an LFC line is swept across a pixel gives a tight estimate of the proportionality coefficient $ c/(G\cdot P)$ . Note that this coefficient captures in it the effective width and $q_d$ of the sub-pixel defect, even though our model was a naive step function.

\begin{figure}
\begin{center}
\begin{tabular}{c c}
\includegraphics[width=0.4\textwidth]{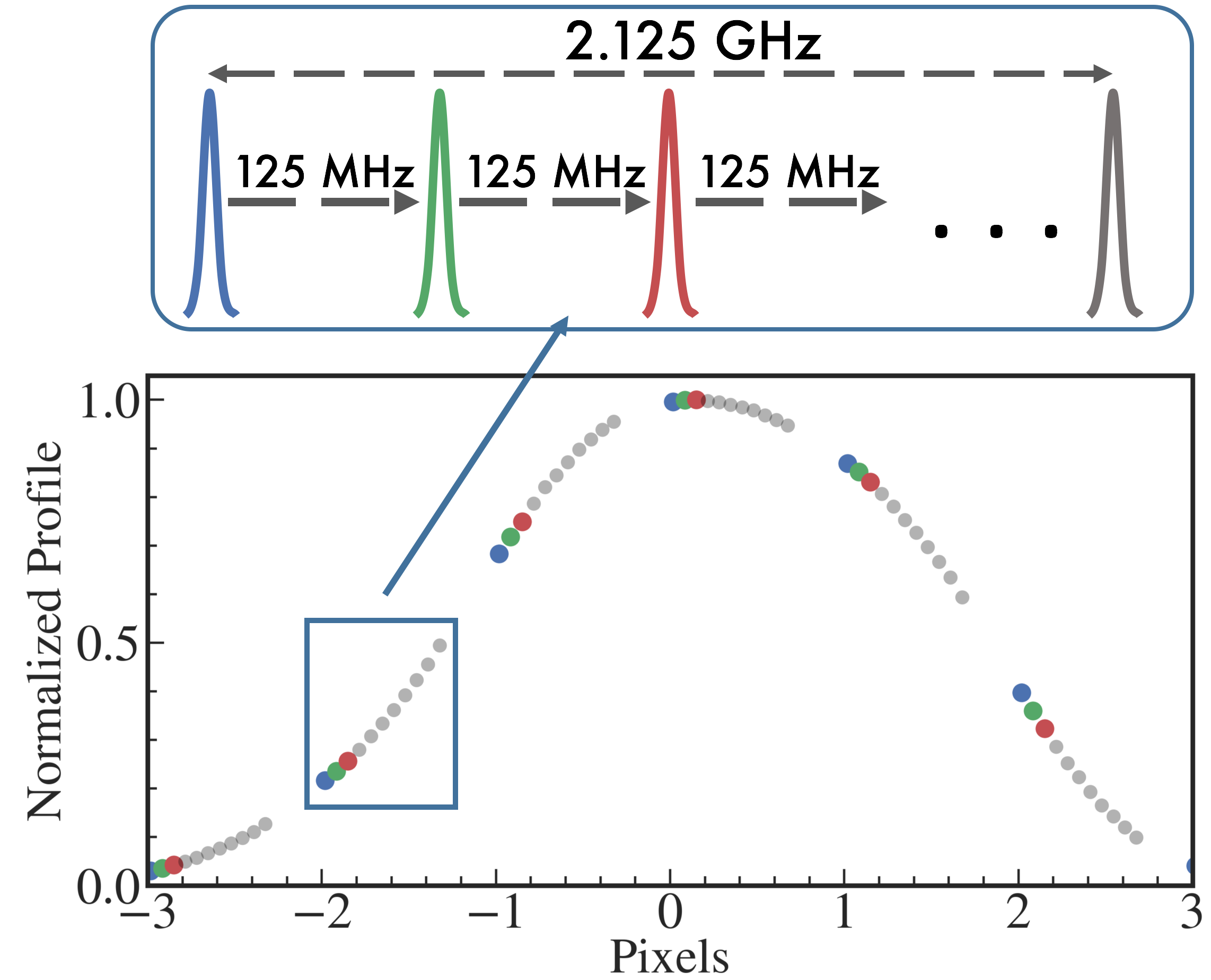} & \includegraphics[width=0.3\textwidth]{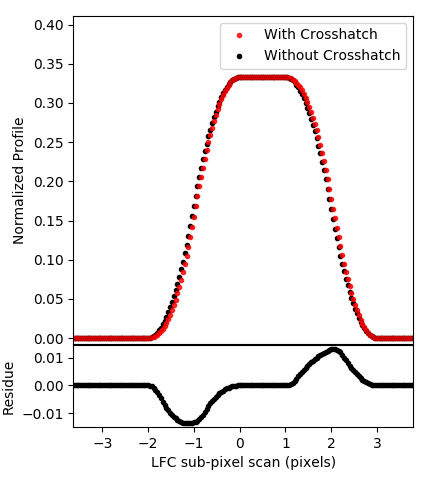}\\
(a) & (b) \\
\end{tabular}
\end{center}
\caption 
{ \label{fig:LFCScan}
a) Current capability of the HPF LFC: Experimental results showing the 1pixel scanning (in 125 MHz steps) of the HPF's PSF. b) Red curve shows the simulated LFC profile traced by a pixel with intra-pixel QE defect, during a full profile scan with the future upgraded LFC. Black curve is the reference curve when all the pixels are free of crosshatch defects. } 
\end{figure}

Currently HPF's LFC\cite{Metcalf2019} enables us to scan only a frequency range corresponding to a 1 pixel width in the HPF focal plane. Figure \ref{fig:LFCScan} (a) shows the super-resolution trace of the instrumental profile we generated by scanning the current LFC at sub-pixel positions. After a proposed future upgrade to the HPF LFC, we will be able to scan the LFC by 30 GHz, enabling us to scan all the pixels in the spectrum with sub-pixel resolution. Figure \ref{fig:LFCScan} (b), shows a simulation of the normalised profile trace we would obtain for a simulated single pixel with intra-pixel QE defect. The difference in the profile to the normal profile of the LFC as a function of slope will directly constrain the correction coefficient.

\section{Conclusion}
\label{sect:conclusion}
Using a simple toy-model we have shown that the cross-hatch pattern induced correction term is additive and proportional to the slope of the underlying flux illumination on the pixel. Our simple model also enables calculation of the RV impact of these cross-hatch patterns. We show that spreading the light across multiple pixels in the cross-dispersion helps ameliorate the impact of these defects, which we estimate to be $\sim$0.4 m/s for HPF (with stated caveats, including the assumption that most or all of the pixels in the order suffer from cross-hatch issues) for RVs from a single order of a typical mid M-star. {\bf  In reality, for the actual HPF detector, the red halfs of the red-most orders are significantly more affected by this issue (k$\sim$1) than the blue halfs of the blue-most orders (k$\sim$0.1).} Future work will model the actual HPF detector. We also present a technique to determine the typical characteristics of these sub-pixel cross-hatch patterns from flat images. We find that for HPF's detector the typical width of the crosshatch pattern is $\sim 5$ $\mu m$, and the QE inside the defect relative to outside is $\sim$54 \%. The QE varies significantly across the defects, and the density of the cross-hatches also varies significantly across the detector.
Modern NIR detectors still remain as a major source of systematic error in precision NIR RVs. While better detectors are always the solution, we also demonstrate a proof-of-concept experiment to enable characterization of these defects employing a tunable frequency stabilized laser comb (LFC).

%In this manuscript using a simple step model, we have shown the correction term in spectrum due to cross-hatch pattern is additive in nature and proportional to the slope of the underlying flux illumination on the pixel. We also provide a simple model to calculate the impact on RV for NIR precision RV spectrographs like HPF. The spreading of spectrum across multiple pixels significantly helps to reduce the impact of crosshatch on the RV, provided crosshatches are not perfectly aligned in cross dispersion direction. For HPF we estimate the impact to be around 0.4 m/sec for RV form a single order of an M-type star.
%We also present a method to measure the typical characteristics of the sub-pixel crosshatch pattern like width, QE inside the defect and angle, etc from flat images. For HPF's H2RG detector, the typical width is $\sim 5$ $\mu m$, and the QE inside the defect relative to outside is $\sim$ 54\%. The QE varies across the defects. The density of the crosshatches also varies significantly across the detector.
%In the end, we also present our proposed correction algorithm based on the presented model. The upgraded LFC will enable HPF to calibrate the coefficients required for correction.

% \disclosures 
\subsection*{Disclosures}
The authors declare that there is no known conflict of interest. The mention of specific companies and trade names is for scientific and technical purposes only and does not constitute and endorsement by NIST.

\acknowledgments 
Authors thank organizers of ISPA-2018 for organizing a very informative and interactive workshop.
This work was partially supported by the funding from The Center for Exoplanet and Habitable Worlds. The Center for Exoplanet and Habitable Worlds is supported by The Pennsylvania State University, The Eberly College of Science, and The Pennsylvania Space Grant Consortium. We acknowledge support from NSF grants, AST1006676, AST1126413, AST1310885, AST 1310875, and the NASA Astrobiology Institute (NNA09DA76A) in our pursuit of precision radial velocities in NIR, and support from the Heising-Simons Foundation. Computations for this research were performed on the Pennsylvania State University's Institute for CyberScience Advanced CyberInfrastructure (ICS-ACI).  We thank the HET staff for their critical assistance, expertise and support. Data presented herein were obtained at the Hobby-Eberly Telescope
(HET), a joint project of the University of Texas at Austin, the Pennsylvania State University,
Ludwig-Maximilians-Universität München, and Georg-August Universität Gottingen. The
HET is named in honor of its principal benefactors, William P. Hobby and Robert E. Eberly. The HET collaboration acknowledges
the support and resources from the Texas Advanced Computing Center.

Software: astropy\cite{Astropy2018}, numpy\cite{Numpy2011}, scipy\cite{Scipy2001}, matplotlib\cite{Hunter2007}, CoCalc\cite{sage}, Sympy\cite{SymPy}

%%%%% References %%%%%

\bibliography{report}   % bibliography data in report.bib
\bibliographystyle{spiejour}   % makes bibtex use spiejour.bst

%%%%% Biographies of authors %%%%%

%\vspace{2ex}\noindent\textbf{First Author} is an assistant professor at the University of Optical Engineering. He received his BS and MS degrees in physics from the University of Optics in 1985 and 1987, respectively, and his PhD degree in optics from the Institute of Technology in 1991.  He is the author of more than 50 journal papers and has written three book chapters. His current research interests include optical interconnects, holography, and optoelectronic systems. He is a member of SPIE.

%\vspace{1ex}
%\noindent Biographies and photographs of the other authors are not available.

\listoffigures
\listoftables

\end{spacing}
\end{document}